\documentclass[runningheads]{llncs}
\usepackage{graphicx}
\usepackage{comment}
\usepackage{amsmath,amssymb} % define this before the line numbering.
\usepackage{color}

\usepackage{bbm}
\usepackage{bm}

\usepackage[utf8]{inputenc} % allow utf-8 input
\usepackage{url}            % simple URL typesetting
\usepackage{booktabs}       % professional-quality tables
\usepackage{amsfonts}       % blackboard math symbols
\usepackage{nicefrac}       % compact symbols for 1/2, etc.
\usepackage{microtype}      % microtypography
\usepackage{color}
\usepackage[labelformat=empty]{subfig}
\usepackage{booktabs}
\usepackage{array}
\usepackage{multirow}
\usepackage{mathtools}
\usepackage{stmaryrd}
\usepackage[para,online,flushleft]{threeparttable}
\usepackage{float}
\usepackage{comment}
\usepackage{amsmath}
\usepackage{svg}
\usepackage{xcolor}

\newcolumntype{P}[1]{>{\centering\arraybackslash}p{#1}}

\usepackage[linesnumbered,ruled,vlined]{algorithm2e}

\SetCommentSty{mycommfont}
\SetKwInput{KwInput}{Input}                % Set the Input
\SetKwInput{KwOutput}{Output}

\newcommand{\bx}{{\mathbf{x}}}

\newcommand{\bbR}{{\mathbb{R}}}

\newcommand{\cL}{{\mathcal{L}}}

\makeatletter
\DeclareRobustCommand\onedot{\futurelet\@let@token\@onedot}
\def\@onedot{\ifx\@let@token.\else.\null\fi\xspace}

\def\ie{\emph{i.e}\onedot}

\makeatother

\newcommand*\samethanks[1][\value{footnote}]{\footnotemark[#1]}

\makeatletter
\renewcommand{\paragraph}{%
  \@startsection{paragraph}{4}%
  {\z@}{0.5\baselineskip \@plus 0ex \@minus 0ex}{-1em}%
  {\normalfont\normalsize\bfseries}%
}
\makeatother
\begin{document}

\pagestyle{headings}
\mainmatter
\title{A Cascaded Learning Strategy for Robust COVID-19 Pneumonia Chest X-Ray Screening}

% Chest X-Ray notation refer to https://www.nhlbi.nih.gov/health-topics/chest-x-ray
%
\titlerunning{A Cascaded Learning Strategy for Robust COVID-19 Screening}
\author{Chun-Fu Yeh\thanks{Both authors contributed equally to this work.}\inst{1} \and
        Hsien-Tzu Cheng\samethanks\inst{1}\and
        Andy Wei\inst{1} \and
        Hsin-Ming Chen\inst{3} \and \\
        Po-Chen Kuo\inst{1} \and
        Keng-Chi Liu\inst{1}\and
        Mong-Chi Ko\inst{1} \and
        Ray-Jade Chen\inst{5} \and \\
        Po-Chang Lee\inst{6} \and
        Jen-Hsiang Chuang\inst{7} \and
        Chi-Mai Chen\inst{8} \and
        Yi-Chang Chen\inst{3} \and \\
        Wen-Jeng Lee\inst{3,4} \and
        Ning Chien\inst{3} \and
        Jo-Yu Chen\inst{3} \and
        Yu-Sen Huang\inst{3,4} \and \\
        Yu-Chien Chang\inst{3} \and
        Yu-Cheng Huang\inst{3} \and
        Nai-Kuan Chou\inst{2} \and
        Kuan-Hua Chao\inst{1} \and \\
        Yi-Chin Tu\inst{1} \and
        Yeun-Chung Chang\inst{3,4}\thanks{Corresponding authors. {Email: {\tt  contact@taimedimg.tw, ycc5566@ntu.edu.tw}}} \and
        Tyng-Luh Liu\inst{1}\samethanks
}
\authorrunning{C. Yeh et al.}
\institute{Taiwan AI Labs \\
\and Division of Cardiovascular Surgery, National Taiwan University Hospital \\
\and Department of Medical Imaging, National Taiwan University Hospital \\
\and Department of Radiology, National Taiwan University College of Medicine \\
\and Department of Surgery, Taipei Medical University Hospital \\
\and National Health Insurance Administration, Ministry of Health and Welfare, Taiwan \\
\and Centers for Disease Control, Ministry of Health and Welfare, Taiwan \\
\and Taiwan Executive Yuan\\
}
%\end{comment}
%******************
\maketitle

\begin{abstract}
We introduce a comprehensive screening platform for the COVID-19 (a.k.a., SARS-CoV-2) pneumonia. The proposed AI-based system works on chest x-ray (CXR) images to predict whether a patient is infected with the COVID-19 disease. Although the recent international joint effort on making the availability of all sorts of open data, the public collection of CXR images is still relatively small for reliably training a deep neural network (DNN) to carry out COVID-19 prediction. To better address such inefficiency, we design a cascaded learning strategy to improve both the sensitivity and the specificity of the resulting DNN classification model. Our approach leverages a large CXR image dataset of non-COVID-19 pneumonia to generalize the original well-trained classification model via a cascaded learning scheme. The resulting screening system is shown to achieve good classification performance on the expanded dataset, including those newly added COVID-19 CXR images. More specifically, the proposed DNN learning proceeds in three stages. In the first stage, the model is trained to predict the mask of lung regions to emphasize the targeted areas of concern and alleviate the effect of irrelevant annotations on a CXR image. Incremental learning is then deployed in the subsequent stage so that the pre-trained DNN can learn to classify the additional COVID-19 images, while retaining its classification performance on the original data. The design is to first filter out images of the normal category in the second stage and then, in the final stage, conduct fine-grained classification to divide the pneumonia candidates into two specific types, namely, COVID-19 and non-COVID-19. We report promising results on both the open and clinical COVID-19 datasets. The proposed method has been integrated into the COVID-19 screening system maintained by Taiwan Centers for Disease Control and Taiwan National Health Insurance Administration. The screening platform is available for testing at \url{https://covirus.cc/pneumonia}.
%with guest account (Username: guest, Password: guest). % Insightful discussions and scientific findings about the proposed AI-based model are also provided in detail.
\end{abstract}

% keywords can be removed

%
\section{Introduction}
The outbreak of COVID-19 disease (a.k.a., SARS-CoV-2) has been affecting the world in an unprecedented way. While intensively global research efforts are being made to seek its effective treatments or vaccines, at the core of the most urgent concern is to prevent the pandemic from further spreading into an uncontrollable and chaotic status. In this work, we are endeavoring to establish an open AI-based platform to seamlessly carry out preliminary large-scale screening of potential COVID-19 patients, and provide early detection at the onset of the infection from the viewpoint of radiology imaging.

Countries such as the United States and Japan are now actively practicing social distancing and seem to receive encouraging outcomes to curb further spread of COVID-19. Still, an affordable and reliable procedure to effectively screen the potentially infected patients is much needed. The reverse-transcription polymerase chain reaction (RT-PCR) testing is the preliminary evaluation to assess whether a subject is at risk of COVID-19. However, the less satisfactory sensitivity of RT-PCR has its limitation on reducing the false-negative rate \cite{fang2020sensitivity} and may overlook a good number of  COVID-19 patients, especially in their early stage of infection. It is essential to find a complementary testing to boost the screening confidence and chest radiology, especially chest x-ray (CXR), with the advantages of affordability, efficiency and reliability, is promising in this regard. As will be demonstrated in the experiments, the proposed AI-based CXR screening system of COVID-19 could effectively detect the infection with high sensitivity, even several days prior to the confirmed RT-PCR results.

Compared with chest computer tomography (CT), chest x-ray imaging is more suitable for being incorporating into a large-scale AI-based screening platform for COVID-19. The supporting evidence is threefold. First, performing CT scanning for comprehensive screening is not practical to contain a pandemic outbreak in that even the well-established healthcare system of a developed country simply does not have such a capacity to do so. Second, cleaning the CT scanning equipment takes considerably longer time than the case of using chest x-ray. The deep cleaning is necessary to prevent subsequent patients from being infected of COVID-19. Third, chest x-rays are already widely adopted as a de facto screening procedure, while CT scanning equipment is not equally popular and is mostly available in primary healthcare institutes. Their distinction in public use also reflects in the availability of large-scale open datasets of chest x-ray images over CT. The ease of collecting image data is crucial in training a reliable AI-based screening system, especially under the current circumstance.

The proposed chest x-ray screening platform leverages the pneumonia classification system developed by Taiwan AI Labs to detect COVID-19. To extend the model, we collaborate with several medical research centers in Taiwan to collect chest x-ray images from COVID-19 patients at various stages, and re-train the pneumonia classification system using a three-stage cascaded learning strategy. Specifically, in the first stage, the AI model is trained to predict from a given CXR image the regions of interest, \ie, the mask of lung regions. In the second stage, features from the predicted regions are extracted to decide whether the image is a positive case of pneumonia. In the final and the third stage, the platform would further make decision on whether the underlying case is COVID-19 or another type of pneumonia. The resulting screening system also yields the stage-wise predicted heatmaps and thus provides explainable image clues leading to the classifications. To demonstrate the effectiveness of the proposed platform for COVID-19 screening, we report the inference results from both the open datasets and those from the collected clinical cases.

\section{Methodology}
\label{sec:method}

% Figure for DNN architecure
\begin{figure*}[!t]
\begin{center}
\includegraphics[width=0.95\linewidth]{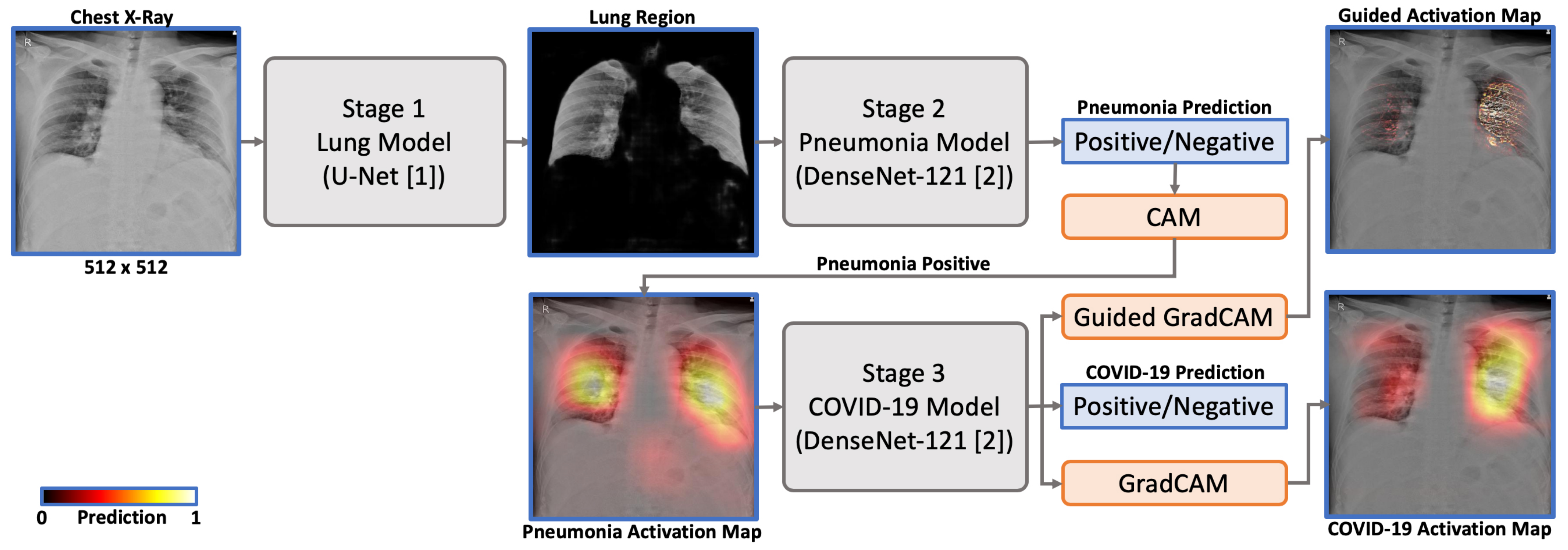}
%\vspace{-3mm}
\caption{Our AI-based COVID-19 screening platform.} % The CXR imaging model explores a 3-stage cascaded learning by focusing on the lung mask (stage 1) and then the pneumonia-related regions (stage 2) to yield the heatmap, indicating the most relevant areas to the predicted probability of COVID-19 infection.}
\label{fig:architecture}
\end{center}
\end{figure*}

We denote the dataset for learning the classification network as $D = \{(\bx_i, y_i)\}_{i=1}^N$, where $\bx_i$ is a chest x-ray image and $y_i$ is the pneumonia class label. In our implementation, we have $y_i = 0$ to reflect that $\bx_i$ is a CXR image of {\tt Normal} case and $y_i \in \{1,2\}$ for the {\tt COVID-19} and {\tt non-COVID-19} Pneumonia, respectively. The training set can be further decomposed as $D = D_o \cup D_c$ and $|D_o| \gg |D_c|$ to indicate that $D_o$ is the original (large) collection for conventional pneumonia classification and $D_c$ includes all the COVID-19 CXR images.  The overall network architecture of the proposed three-stage cascaded learning is illustrated in Fig.~\ref{fig:architecture}. We detail the design details as follows.

\begin{figure*}[!t]
\begin{center}
\includegraphics[width=0.95\linewidth]{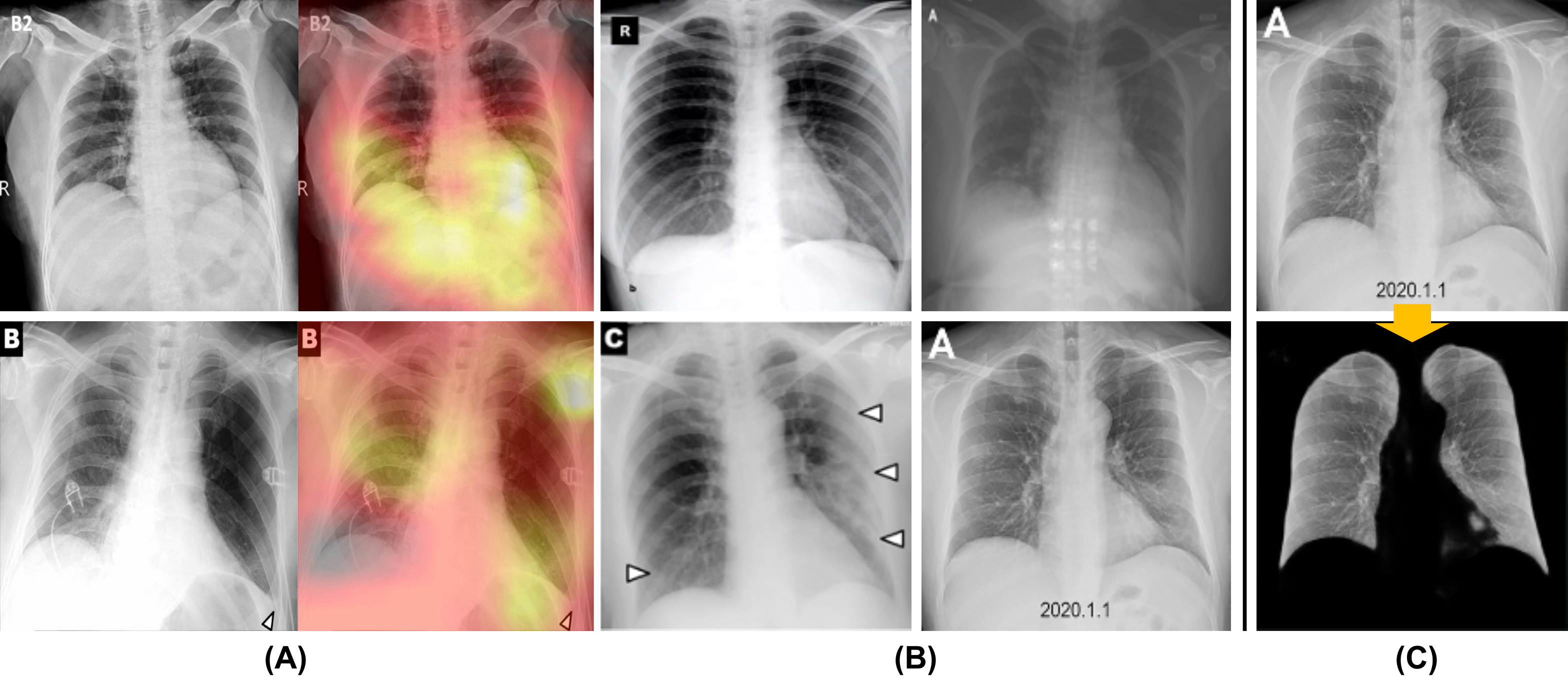}
% \includegraphics[width=0.90\linewidth]{fig/sampleswithnoises.pdf}
%\vspace{-3mm}
\caption{(A) Misleading model attentions (right column) caused by non-informative areas (left column). (B) Four examples of ``noisy'' CXR images from open datasets. The alphabet at the upper-right corner, date stamp, and arrow marks in CXR images may confuse an AI-based system in both training and inference. (C) Stage 1 of the proposed DNN architecture functions as a preprocessing step to filter out non-informative areas and enable the system to focus on the lung regions.}
\label{fig:noisy_samples}
\end{center}
\end{figure*}

\paragraph{Lung segmentation.}
The convenience of working with an open dataset often comes with a price---the image quality could vary among samples in the collection and noisy or irrelevant annotations may be included in some images. Figures~\ref{fig:noisy_samples}(A) and \ref{fig:noisy_samples}(B) illustrate examples of such issues regarding the image or annotation quality from an open dataset of medical images.

To prevent an AI-based system from learning inconsistent information or noisy annotations, we design the pneumonia classification system to focus on the essential regions of interest, which in our case is the lung areas. We consider a U-Net model \cite{unet2015} to first predict a mask of the lung regions in a CXR image, and use the resulting mask to filter out image areas that could distract the AI system from learning the crucial pneumonia-relevant features for the underlying classification problem. Simply put, the task of the first stage, as shown in Fig.~\ref{fig:noisy_samples}(C), can be thought of as a preprocessing step to segment the mask of the lung regions in each given CXR image.

% Figure for samples with noise

% A paragraph to describe the U-Net model, including the U_Net architecture, and the making of GT

\paragraph{Pneumonia classification.}
Having excluded non-informative regions from each CXR image $\bx_i$, the task of the succeeding second stage of the proposed cascaded learning is to carry out binary classifications: {\tt Normal} versus {\tt Pneumonia}. That is, the DenseNet-121 \cite{densenet2017} pneumonia classifier is trained to predict $\bx_i$ as negative if $y_i =0$, and otherwise positive if $y_i \in \{1, 2 \}$. Under this design setting, all CXR images of pneumonia are expected to be classified as positive no matter what their type is. After all, the goal of this stage is to filter out non-pneumonia samples from further considerations.

% Need a figure to illustrate the backbone and the stage 2 details

It is constructive to explain the pneumonia classification outcome of a CXR image $\bx_i$ by using the CAM \cite{zhou2016learning} interpretation technique. As shown in Fig.~\ref{fig:architecture}, performing global average pooling (GAP) yields a $1024$-D feature vector $f^2_i$, which can be reshaped into a $32\times 32$ tensor $h^2_i$ and then upsampled to $H^2_i$ of the original input resolution, $512\times 512$. We express the CAM modeling at stage 2 as follows.
\begin{equation}
    f^2_i \in \bbR^{1024} \; \xmapsto[]{reshape} \; h^2_i \in \bbR^{32\times 32} \; \xmapsto[]{upsample} \; H^2_i \in \bbR^{512\times 512}.
    \label{eqn:cam2}
\end{equation}

The heatmap $H^2_i$ derived in (\ref{eqn:cam2}) can be interpreted as the importance distribution over the lung regions leading to the pneumonia classification outcome of $\bx_i$. Intuitively, the AI-based readout of a CAM heatmap can be used to aide physicians to identify anomalous spots in the input CXR image $\bx_i$.

Recall that the training dataset $D$ comprises $D_o$ and $D_c$. The latter includes the COVID-19 images and is expected to expand constantly when new CXR samples are provided by our collaborative medical centers. The frequent changes in the available training data may require re-training the AI model to improve classification performance on COVID-19 images, while it also may degrade the overall pneumonia classification in stage 2. To overcome this dilemma, we deploy incremental learning to ensure a robust model learning in stage 2. We first adopt a pneumonia classification model by Taiwan AI Labs, which is pre-trained on $D_o$, and then perform incremental learning on $D_c$ by optimizing the model parameters $\theta$ with respect to the following loss function:
\begin{equation}
    \cL_D(\theta) = \frac{1}{|D|} \left(\,\sum_{\bx_i \in D} \ell_{\mathrm{CE}}(\bx_i,y_i) + \lambda\,\sum_{\bx_i \in D_o} \ell_{\mathrm{KL}}(\bx_i)\,\right)
    \label{eqn:loss}
\end{equation}
\noindent where $\ell_{\mathrm{CE}}$ is the cross-entropy loss, $\ell_{\mathrm{KL}}$ is the knowledge distillation loss for learning the prediction output of the pre-trained model and $\lambda \geq 0$ is a parameter to weigh the effect of knowledge distillation. When $\lambda$ is set to $0$, the incremental learning is simply reduced to re-training the model with the updated data $D$.

\paragraph{COVID-19 screening.}
At the stage 3 of the cascaded learning, the task is to distinguish the specific type of pneumonia: {\tt COVID} ($y=1$) and {\tt non-COVID} ($y=2$). Notice that the dataset $D_o$ solely comprises non-COVID CXR images, which could correspond to normal cases and viral or bacterial pneumonia.
\begin{comment}
Although our goal is to detect those CXR images with high possibility of COVID-19 infection, it is insightful to consider fine-grained pneumonia classification over three possible categories, namely, {\tt COVID} ($y=1$), {\tt bacteria} ($y=2$) and {\tt viral} ($y=3$) pneumonia. This way we can be confident that the AI-based system can capture not only the distinctions between viral and bacteria pneumonia, but also the subtle differences between COVID-19 and other viral pneumonia.
\end{comment}
We can write the input to the stage 3 classification as
\begin{equation}
    \tilde{\bx}_i = \bx_i \odot H^2_i
    \label{eqn:input3}
\end{equation}
\noindent where $\odot$ symbolizes pixel-wise product. In Fig.~\ref{fig:architecture}, an example of $\tilde{\bx}_i$ is illustrated as a mask of informative regions relevant to predicting an input CXR image $\bx_i$ as a {\tt COVID-19} or {\tt non-COVID-19} pneumonia case at the stage 3.

Analogous to the derivation of $H^2_i$ in (\ref{eqn:cam2}), we can explore the heatmap responses to gain insights into how the classification system distinguishes COVID-19 from other types of pneumonia. We consider GradCAM \cite{selvaraju2017grad} to generate the heatmap $H^3_i$ in that the resulting GradCAM responses tend to be concentrated rather than diffusive. The heatmap $H^3_i$ yields interpretable image cues relevant to how likely the AI system predicts the pneumonia case of $\bx_i$ as COVID-19. Inspired by \cite{selvaraju2017grad}, we also generate Guided-GradCAM (Guided Activation) as detailed responses to further focus on patterns to distinguish COVID-19 pneumonia.

%scientific findings
%
\section{Experiments}
\label{sec:result}

\begin{table}[h!] %[htbp]
\caption{Open datasets used in the experiments. Abbreviations: N: {\tt Normal}, P: Pneumonia ({\tt non-COVID-19}), C: {\tt COVID-19} \cite{cohen2020covid}}
  \centering
    \small
    % \begin{tabular}{|l|p{1.5cm}|p{1cm}|p{1cm}|p{1cm}|p{1cm}|p{1cm}|p{1cm}|}
    \begin{tabular}{|P{3.7cm}|P{1cm}|P{1cm}|P{1cm}|P{0.8cm}|P{0.8cm}|P{1.0cm}|P{1.0cm}|P{0.8cm}|}
    \hline
    \multirow{2}{*}{Data source} & \multicolumn{4}{c|}{Number of images} & \multicolumn{4}{c|}{Data split (train/val/test)} \\
    \cline{2-9}
    & Total & N & P & C & Total & N & P & C \\
    \hline
    \hline
    \multirow{3}{*}{Padchest} & \multirow{3}{*}{41,364} & \multirow{3}{*}{36,142} & \multirow{3}{*}{5,222} & \multirow{3}{*}{-} & & 28,955 & 4,142 &  \multirow{3}{*}{-} \\
    & & & & & & 3,591 & 547 & \\
    & & & & & & 3,596 & 533 & \\
    \hline
    \multirow{3}{*}{RSNA} & \multirow{3}{*}{18,406} & \multirow{3}{*}{8,851} & \multirow{3}{*}{9,555} & \multirow{3}{*}{-} & & 7,092 & 7,616 & \multirow{3}{*}{-}  \\
    & & & & & & 882 & 947 & \\
    & & & & & & 877 & 992 & \\
    \hline
    \multirow{3}{*}{Covid-chestxray-dataset} & \multirow{3}{*}{167} & \multirow{3}{*}{-} & \multirow{3}{*}{-} & \multirow{3}{*}{167} & 89 & \multirow{3}{*}{-} & \multirow{3}{*}{-} & 89  \\
    & & & & & 43 & & & 43 \\
    & & & & & 35 & & & 35 \\
    \hline
    \end{tabular}%
  \label{tab.open-dataset}%
  % \vspace{-3mm}
\end{table}%

\begin{table}[h!] %[htbp]
\caption{Clinical datasets used in the experiments. Abbreviations: N: {\tt Normal}, P: Pneumonia ({\tt non-COVID-19}), C: {\tt COVID-19}}
  \centering
    \small
    % \begin{tabular}{|l|p{1.5cm}|p{1cm}|p{1cm}|p{1cm}|p{1cm}|p{1cm}|p{1cm}|}
    \begin{tabular}{|P{1.9cm}|P{1.2cm}|P{1.2cm}|P{1.2cm}|P{1.2cm}|P{1.1cm}|P{1.1cm}|P{1.1cm}|P{1.1cm}|}
    \hline
    \multirow{2}{*}{Data source} & \multicolumn{4}{c|}{Number of images/cases} & \multicolumn{4}{c|}{Data split (train/val/test)} \\
    \cline{2-9}
    & Total & N & P & C & Total & N & P & C \\
    \hline
    \hline
    \multirow{3}{*}{NTUH} & \multirow{3}{*}{107/72} & \multirow{3}{*}{37/37} & \multirow{3}{*}{60/25} & \multirow{3}{*}{10/10}   & 48 & 14 & 29 & 5 \\
    & & & & & 30 & 12 & 15 & 3\\
    & & & & & 29 & 11 & 16 & 2\\
    \hline
    \multirow{3}{*}{TMUH} & \multirow{3}{*}{27/8} & \multirow{3}{*}{-} & \multirow{3}{*}{-} & \multirow{3}{*}{27/8} & 16  & \multirow{3}{*}{-} & \multirow{3}{*}{-} & 16 \\
    & & & & & 8 &  &  & 8\\
    & & & & & 3 &  &  & 3\\
    \hline
    \multirow{3}{*}{NHIA} & \multirow{3}{*}{3,375} & \multirow{3}{*}{-} & \multirow{3}{*}{3,069} & \multirow{3}{*}{306}   & 2,033 & - & 1,836 & 197 \\
    & & & & & 668 & - & 614 & 54 \\
    & & & & & 674 & - & 619 & 55\\
    \hline
    \end{tabular}%
  \label{tab.clinical-dataset}%
  % \vspace{-3mm}
\end{table}%

\begin{figure*}[!t]
\begin{center}
\includegraphics[width=0.9\linewidth]{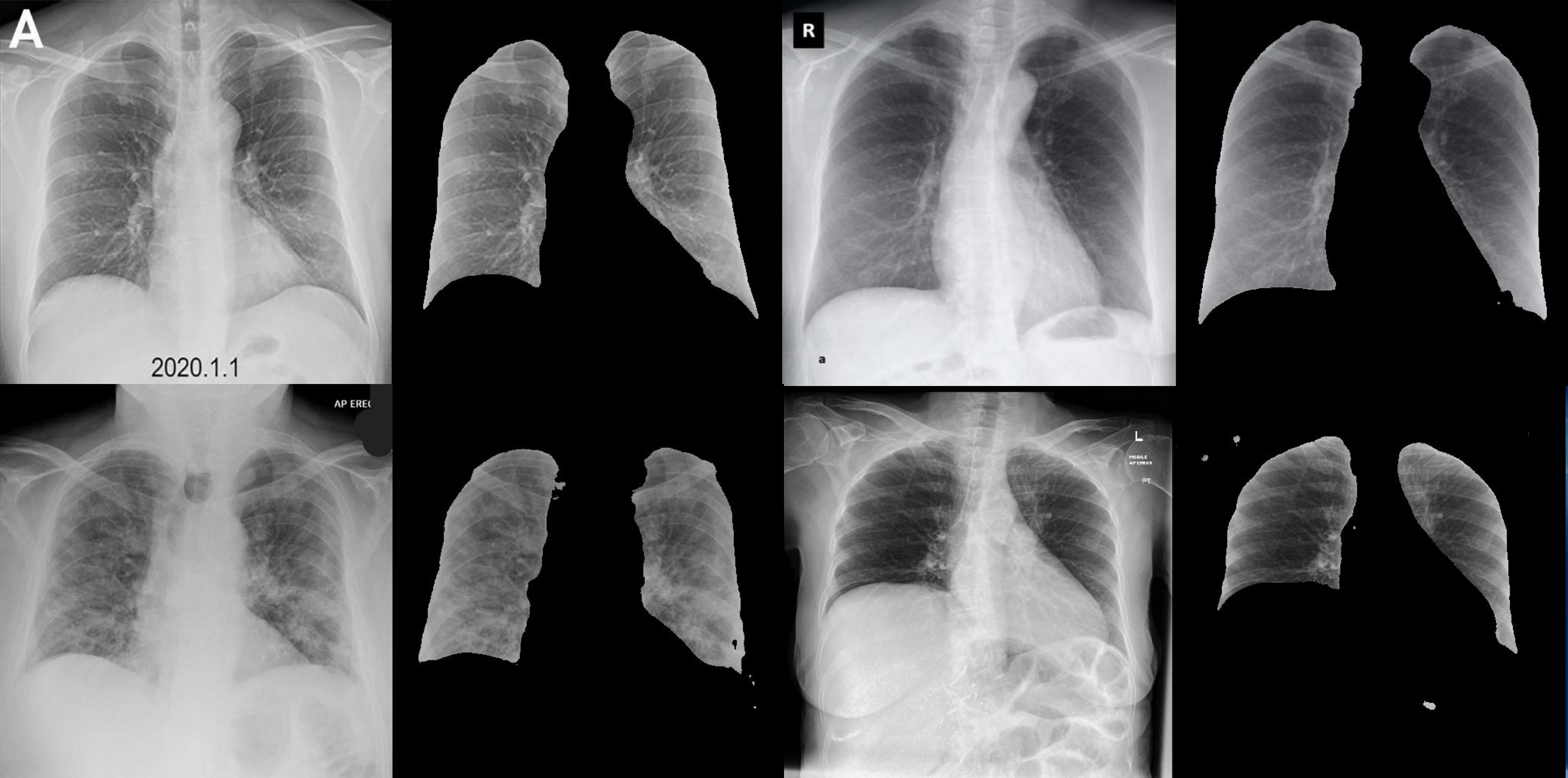}
\caption{Lung segmentation results on Open COVID-19 samples by our stage 1 model.}
\label{fig:lung-seg}
\end{center}
\end{figure*}

In collaborations with NTUH, TMUH and NHIA, we evaluate our method on both open and clinical collections of CXR images. Details about the datasets are listed in Table~\ref{tab.open-dataset} and Table ~\ref{tab.clinical-dataset}. We learn the model of each stage separately, \ie, all parameters other than those of the current training stage are kept unchanged. ~

\paragraph{Lung segmentation masking. }
The preprocessing stage 1, lung segmentation, is trained on Tuberculosis Chest X-ray Image Data Sets that consist of two parts, Montgomery County X-ray Set and Shenzhen Hospital X-ray Set. The total number of samples contains 406 normal case images and 394 tuberculosis case images.  We randomly sample 80\% as our training/validation sets and 20\% as our test set. Our stage 1 model achieves 0.88  dice similarity coefficient (DSC) on the test set. Fig.~\ref{fig:lung-seg} shows some segmentation outcomes from our stage 1 model. We further investigate the effect of stage 1 considering the whole screening pipeline. Table~\ref{tab.stage1-result} compares the pneumonia classification (stage 2) results between models with lung segmentation masking and models without lung segmentation masking.

\begin{table}[t!] %[htbp]
\caption{With ($\checkmark$)/without ($\times$) using stage-1 lung segmentation on open datasets}
  \centering
    \small
    \begin{tabular}{|P{2cm}|P{2cm}|P{2cm}|P{2cm}|P{2cm}|}
    \hline
    Split & Lung mask & AUC & Sensitivity & Specificity \\
    \hline
    \hline
    \multirow{2}{*}{Validation}& $\checkmark$ & 97.58 & 90.64 & 94.95 \\
    \cline{2-5}
    & $\times$ & 97.58 & 90.05 & 95.37 \\
    \hline
    \multirow{2}{*}{Test}& $\checkmark$ & \textbf{97.33} & \textbf{90.37} & \textbf{95.66} \\
    \cline{2-5}
    & $\times$ & 97.18 & 87.56 & 95.44 \\
    \hline
    \end{tabular}%
  \label{tab.stage1-result}%
  % \vspace{-3mm}
\end{table}%

\paragraph{COVID-19 chest x-ray screening using open datasets.}
We first train the stage 2 model to carry out pneumonia classification on the open datasets, summarized in Table~\ref{tab.open-dataset}. We separate the data into three groups: {\tt Noraml} (N), Pneumonia (P, {\tt non-COVID-19}), and {\tt COVID-19} (C). Normal and non-COVID-19 pneumonia CXR images are collected from Padchest \cite{bustos2019padchest} and RSNA \cite{RSNA}, while COVID-19 samples are selected from \cite{cohen2020covid}. For the data from Padchest and RSNA, we randomly divide it into training/validation/testing in the ratio of 80\%/10\%/10\%. As to the data of COVID-19, we split it into training/validation/testing in the ratio of 50\%/25\%/25\%, due to its small sample size. During stage 2, binary labels are considered, where {\tt non-COVID-19} Pneumonia (P) and {\tt COVID-19} (C) samples are labeled as 1 (positive) and {\tt Normal} samples are labeled as 0 (negative). For each batch of training, we evenly sample each group to account for imbalanced distributions among {\tt Normal}, {\tt non-COVID-19} Pneumonia, and {\tt COVID-19} samples. The masked images by stage 1 model (lung segmentation, as in Fig.~\ref{fig:lung-seg}) are used as the input of stage 2 model for both training and inference. After achieving results of pneumonia classification of stage 2 as in Table~\ref{tab.stage23-open}, we then train the stage 3 model to distinguish COVID-19 from non-COVID-19 pneumonia, using only {\tt non-COVID-19} Pneumonia (P) and {\tt COVID-19} (C) training samples. All parameters but those of stage 3 are fixed during training. In this stage, marked as ``{\bf 3}: COVID-19'' in Table~\ref{tab.stage23-open}, {\tt COVID-19} (C) samples are labeled as 1 (positive) while {\tt non-COVID-19} Pneumonia (P) samples are labeled as 0 (negative). Table~\ref{tab.stage23-open} shows the results of our method on both open and clinical datasets, where only open datasets are utilized during the training phase.

\begin{table}[t!] %[htbp]
\caption{Result of our model \textbf{trained on open datasets only}
}
  \centering
    \small
    % \begin{tabular}{|l|p{1.5cm}|p{1cm}|p{1cm}|p{1cm}|p{1cm}|p{1cm}|p{1cm}|}
    \begin{tabular}{|P{2.3cm}|P{1.7cm}|P{2cm}|P{1.7cm}|P{1.7cm}|P{1.7cm}|P{1.7cm}|}
    \hline
    Dataset & Split & Stage & AUC & Sensitivity & Specificity \\
    \hline
    \hline
    \multirow{4}{*}{Open} & \multirow{2}{*}{Validation} & {\bf 2}:Pneumonia & 96.97 & 93.05 & 90.79 \\
    \cline{3-6}
    & & {\bf 3}:COVID-19 & 86.91 & 90.34 & 74.42 \\
    \cline{2-6}
    & \multirow{2}{*}{Test}& {\bf 2}:Pneumonia & 96.72 & 92.85 & 90.05\\
    \cline{3-6}
    & & {\bf 3}:COVID-19 & 88.04 & 85.26 & 85.86\\
    \hline
    \hline
    \multirow{4}{*}{NTUH+TMUH} & \multirow{2}{*}{Validation}& {\bf 2}:Pneumonia & 91.99 & 33.33 & 92.31 \\
    \cline{3-6}
    & & {\bf 3}:COVID-19 & 67.88 & 66.67 & 63.34\\
    \cline{2-6}
    & \multirow{2}{*}{Test}& {\bf 2}:Pneumonia & 93.94 & 63.64 & 90.48 \\
    \cline{3-6}
    & & {\bf 3}:COVID-19 & 40.00 & 50.00 & 40.00\\
    \hline
    \end{tabular}%
  \label{tab.stage23-open}%
  % \vspace{-3mm}
\end{table}%

\begin{table}[t!] %[htbp]
\caption{Result of our model trained on open and clinical datasets
}
  \centering
    \small
    % \begin{tabular}{|l|p{1.5cm}|p{1cm}|p{1cm}|p{1cm}|p{1cm}|p{1cm}|p{1cm}|}
    \begin{tabular}{|P{2.3cm}|P{1.7cm}|P{2cm}|P{1.7cm}|P{1.7cm}|P{1.7cm}|P{1.7cm}|}
    \hline
    Dataset & Split & Stage & AUC & Sensitivity & Specificity \\
    \hline
    \hline
    \multirow{4}{*}{Open} & \multirow{2}{*}{Validation} & {\bf 2}:Pneumonia & 97.00 & 87.07 & 95.57 \\
    \cline{3-6}
    & & {\bf 3}:COVID-19 & 99.70 & 88.37 & 98.90 \\
    \cline{2-6}
    & \multirow{2}{*}{Test}& {\bf 2}:Pneumonia & 96.64 & 86.54 & 95.22 \\
    \cline{3-6}
    & & {\bf 3}:COVID-19 & 99.88 & 91.43 & 99.44\\
    \hline
    \hline
    \multirow{4}{*}{NTUH+TMUH} & \multirow{2}{*}{Validation}& {\bf 2}:Pneumonia & 97.44 & 96.15 & 75.00 \\
    \cline{3-6}
    & & {\bf 3}:COVID-19 & 87.88 & 100.00 & 80.00 \\
    \cline{2-6}
    & \multirow{2}{*}{Test}& {\bf 2}:Pneumonia & 98.70 & 95.24 & 90.91 \\
    \cline{3-6}
    & & {\bf 3}:COVID-19 & 97.50 & 100.00 & 75.00 \\
    \hline
    \end{tabular}%
  \label{tab.stage23-clinical}%
  % \vspace{-3mm}
\end{table}%

\begin{figure*}[!t]
\begin{center}
\includegraphics[width=1.0\linewidth]{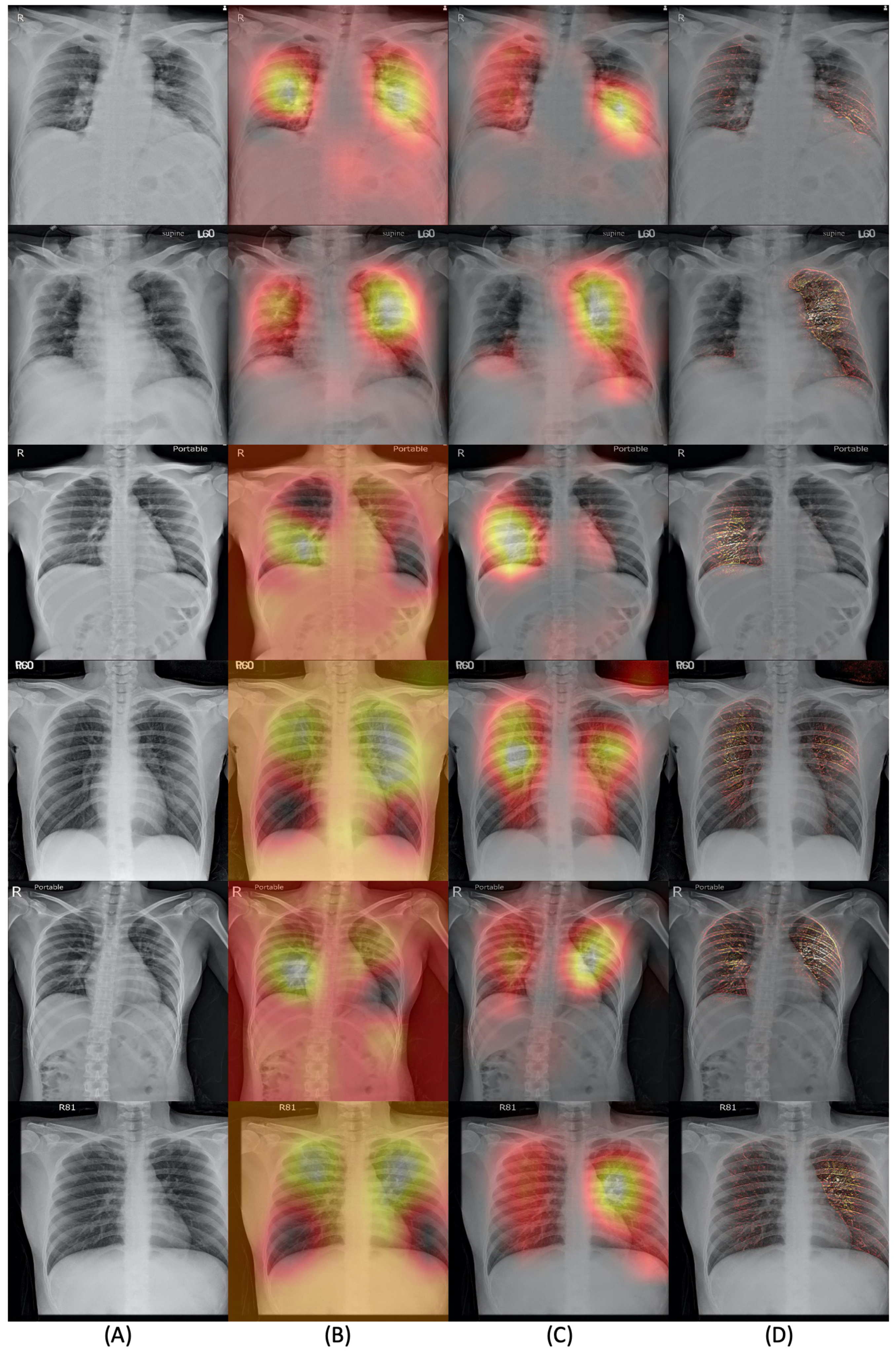}
\caption{Visualization of qualitative results by our method on COVID-19 cases from clinical dataset. (A) Original chest x-ray image. (B) Regions of interest for pneumonia (Stage 2). (C) Regions of interest specific to {\tt COVID-19} (Stage 3). (D) Guided activation specific to {\tt COVID-19} (Stage 3).}
\label{fig:hotmaps_clinical}
\end{center}
\end{figure*}

\begin{figure*}[!t]
\begin{center}
\includegraphics[width=1.0\linewidth]{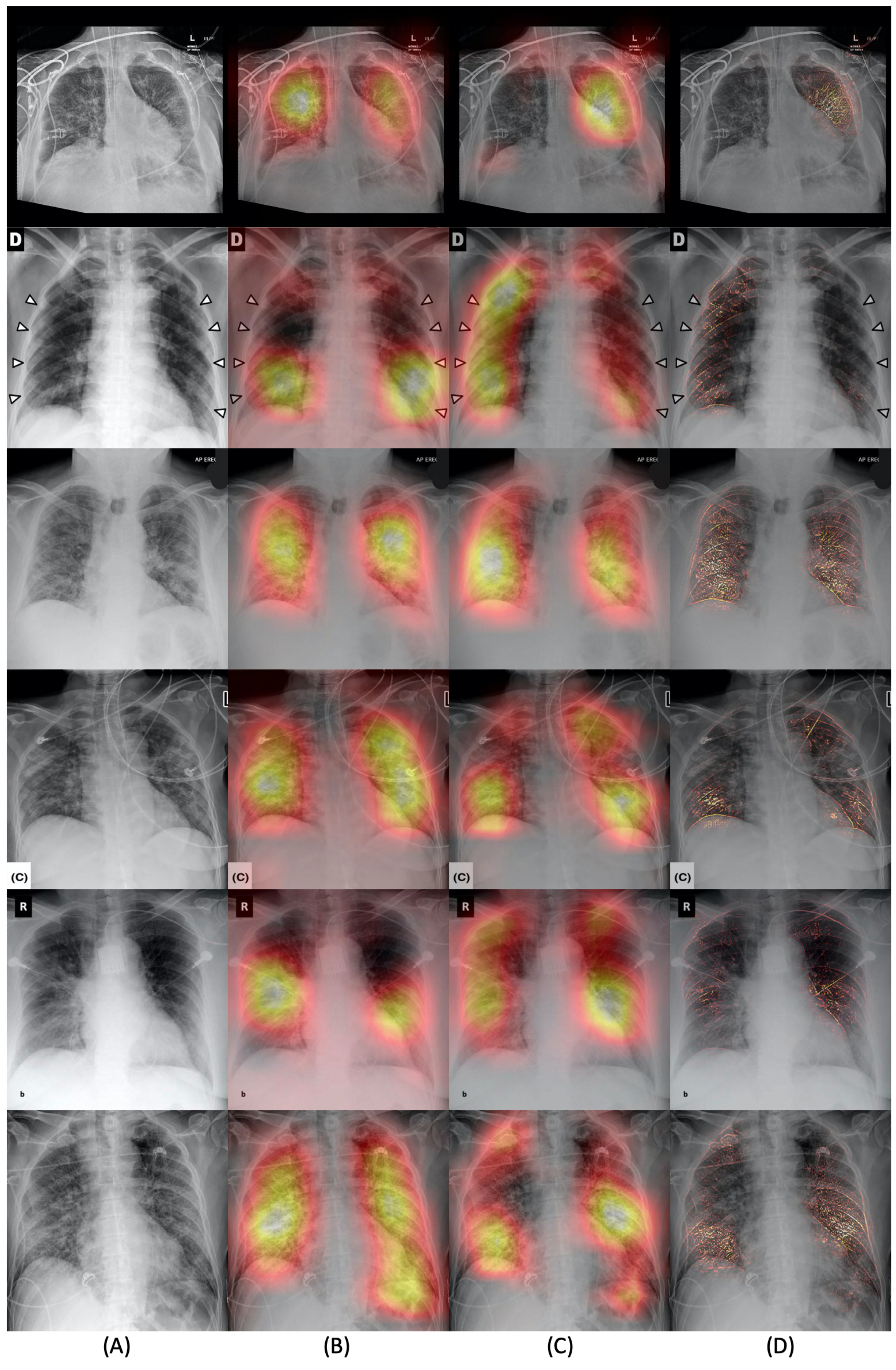}
\caption{Visualization of qualitative results by our method on COVID-19 cases from open dataset. (A) Original chest x-ray image. (B) Regions of interest for pneumonia (Stage 2). (C) Regions of interest specific to {\tt COVID-19} (Stage 3). (D) Guided activation specific to {\tt COVID-19} (Stage 3).}
\label{fig:hotmaps_open}
\end{center}
\end{figure*}

\paragraph{Fine-tune stage 2 and 3 on clinical datasets (NTUH and TMUH).}
From Table~\ref{tab.stage23-open}, we observe the performance gaps in both validation and testing between clinical data and open data. (We will further elaborate this discovery in the next section.) To further improve the AI-based classification performance and achieve clinical applicability, we collaborate with NTUH, TMUH, and NHIA to collect more clinical COVID-19 CXR samples, as summarized in Table~\ref{tab.clinical-dataset}. Starting from a pre-trained model using only open datasets, we fine-tune our model using clinical data. Instead of using only clinical data, we combine open datasets and clinical datasets to prevent from overfitting during our fine-tuning phase, and consider the incremental learning strategy as in (\ref{eqn:loss}). As in the experiments on open datasets, we fine-tune our pneumonia classification (stage 2) and COVID-19 screening (stage 3) sequentially. Based on the final results shown in Table~\ref{tab.stage23-clinical}, our method is able to recognize {\tt Normal}, Pneumonia (both {\tt non-COVID-19} and {\tt COVID-19}), {\tt non-COVID-19}, and {\tt COVID-19} cases respectively with high AUC, sensitivity, and specificity. For qualitative result, those cases sampled from the testing set of open dataset in Fig.~\ref{fig:hotmaps_open} and the clinical dataset in Fig.~\ref{fig:hotmaps_clinical} demonstrate reasonable model activation in different stages.

\begin{table}[t!] %[htbp]
\caption{Result of our stage 3 model trained and evaluated on NHIA dataset}
  \centering
    \small
    % \begin{tabular}{|l|p{1.5cm}|p{1cm}|p{1cm}|p{1cm}|p{1cm}|p{1cm}|p{1cm}|}
    \begin{tabular}{|P{2cm}|P{1.7cm}|P{2cm}|P{1.7cm}|P{1.7cm}|P{1.7cm}|}
    \hline
    Dataset & Split & Stage & Sensitivity & Specificity \\
    \hline
    \hline
    \multirow{2}{*}{NHIA} & \multirow{1}{*}{Validation}& {\bf 3}:COVID-19 & 83.33 & 80.46 \\
    \cline{2-5}
    & \multirow{1}{*}{Test}&
    {\bf 3}:COVID-19 & 81.82 & 80.45 \\
    \hline
    \end{tabular}%
  \label{tab.nhia-result}%
  % \vspace{-3mm}
\end{table}%

\paragraph{Fine-tune stage 3 on NHIA dataset.}
To evaluate the clinical applicability of our method, we fine-tune our stage 3 model on a relatively larger dataset (NHIA), which includes 306 {\tt COVID-19} (C) samples from 306 confirmed cases in Taiwan and 3,069 {\tt non-COVID-19} Pneumonia (P) samples. Of those 3,069 Pneumonia (P) samples, 50\% of them are diagnosed with bacterial pneumonia and the other half are diagnosed with viral pneumonia. Table~\ref{tab.nhia-result} includes the validation and testing results on NHIA dataset, suggesting the potential of using our method for distinguishing COVID-19 from both common bacterial and viral pneumonia.

% Figure for DNN architecure
\begin{figure*}[!t]
\begin{center}
\includegraphics[width=0.96\linewidth]{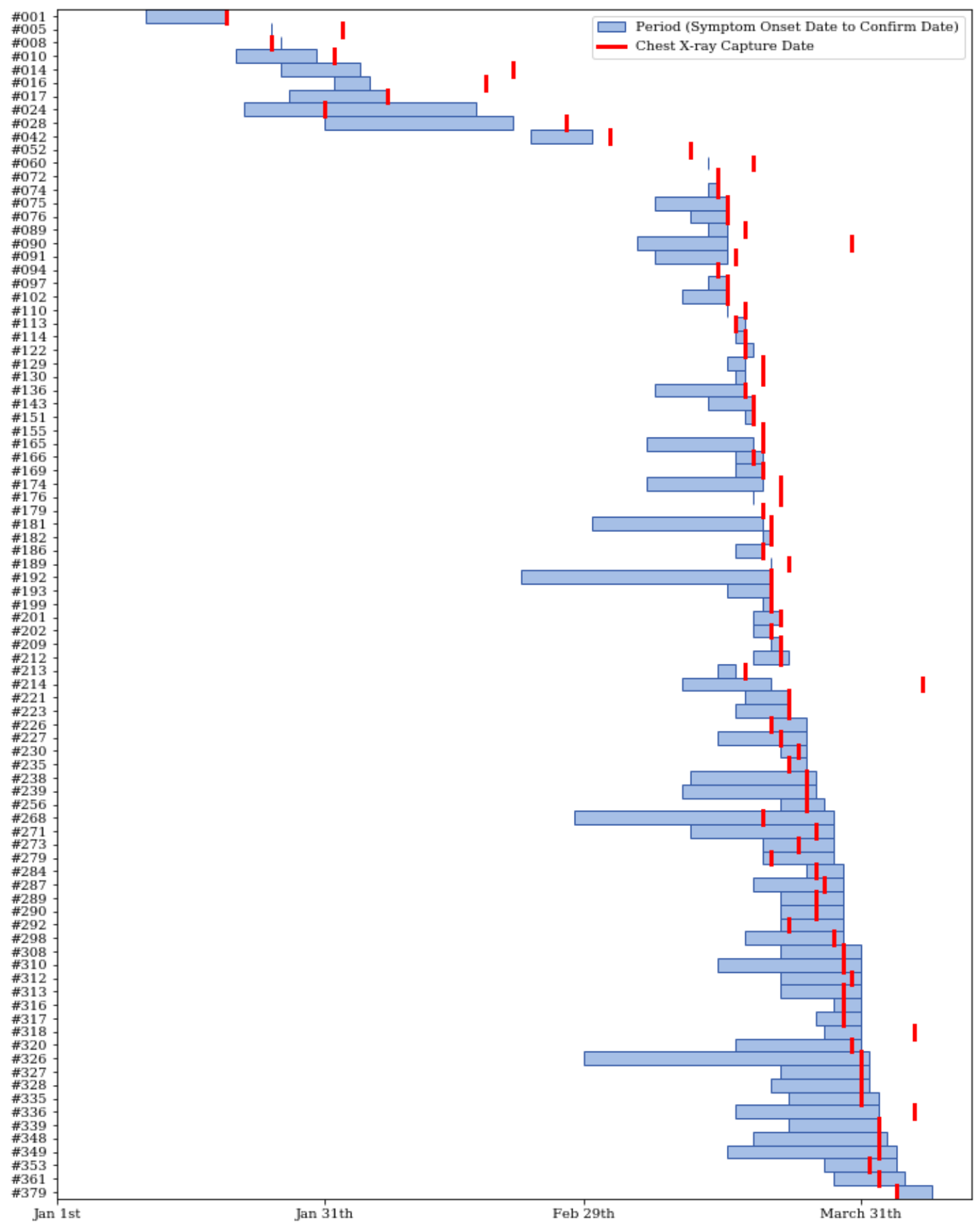}
%\vspace{-3mm}
\caption{The chest x-ray captured date of each case with respect to subjective symptom onset date and confirmed date by RT-PCR. Red line indicates the chest x-ray captured date in which our model also predicted each case as positive (COVID-19). The left border of each light blue box denotes the subjective symptom onset date and the right end indicates the confirmed date. There are only 89 cases out of 109 cases shown in this graph due to missing dates in the other 20 cases.}
\label{fig:caseboxplot}
\end{center}
\end{figure*}

%\paragraph{COVID-19 chest x-ray screening on the NHIA dataset.}
\paragraph{Early COVID-19 detection with our screening system} Further, to gain insights into the advantages of using the proposed CXR screening platform, we investigate its efficiency of early prediction over RT-PCR so that the needed medical treatment can be performed as early as possible. To this end, we describe the overall performance of our screening platform and the case study on the NHIA dataset, particularly the validation (n=54) and the test (n=55) set. Out of these 109 samples detected by RT-PCR, there are 47 samples with visible signs of pneumonia, which are double confirmed by our medical specialist. Based on this subset, our model correctly detects 42 samples; that is, the sensitivity of our model achieves 89.36\%.

Out of these 109 samples, taken in a period of three months (January 2020 to March 2020), the advantage of using our system for early detection of COVID-19 versus using RT-PCR can be observed from in the comparative study. Fig.~\ref{fig:caseboxplot} illustrates the details of each case.
\begin{itemize}
  \item Case 24 was detected by our system, 17 days prior to RT-PCR confirmed (on January 31, 2020).
  \item Case 268 was detected by our system, 6 days prior to RT-PCR confirmed (on March 20, 2020).
  \item There were 27 out of 109 cases detected at least 2 days earlier than the RT-PCR confirmed dates (Case 24, 226, 227, 235, 256, 268, 271, 272, 273, 279, 284, 287, 289, 290, 292, 308, 310, 313, 316, 317, 335, 349, 353, 355, 361, 363, and 379). Among these cases, Case 24, 268, 279, and 292 were detected 5 days earlier.
\end{itemize}

% \begin{figure*}[!t]
% \begin{center}
% \includegraphics[width=1.0\linewidth]{clinicalhotmaps_v3.pdf}
% \caption{The visualization of the performance. (A) Original Chest X-ray Image. (B) Lung Segmentation Result. (C) Region of Interest for Pneumonia-like Lesion. (D) Region of Interest Specific to SARS-CoV-2. (E) Guided Attention Specific to SARS-CoV-2.}
% \label{fig:clinicalhotmaps}
 %\end{center}
% \end{figure*}

\paragraph{Comparison with relevant studies.}
To evaluate the efficacy of our multi-stage method, we further conduct an experiment on the training/testing data split setting of COVID-Net~\cite{wang2020covid}. The CXR samples are taken entirely from open datasets~\cite{cohen2020covid,RSNA}, which consist of 13,594 CXR images (7966/5476/152 in N/P/C) in the training set and 231 CXR images (100/100/31 in N/P/C) in the testing set. As shown in Table~\ref{tab.covidnetcomparison}, our method is more sensitive (and specific at inverse perspective of binary classification) than COVID-Net in both {\tt Normal} and {\tt COVID-19} predictions. Note that in the stage 2 of our method, the binary classification is to predict {\tt Normal} versus both {\tt COVID-19} and {\tt non-COVID-19} pneumonia.

\begin{table}[h!] %[htbp]
\caption{Classification comparison of our method versus  COVID-Net~\cite{wang2020covid}}
  \centering
    \small
    % \begin{tabular}{|l|p{1.5cm}|p{1cm}|p{1cm}|p{1cm}|p{1cm}|p{1cm}|p{1cm}|}
    \begin{tabular}{|P{3.5cm}|P{1.7cm}|P{1.7cm}|P{1.7cm}|}
    \hline
    \multirow{2}{*}{Dataset} & \multicolumn{3}{c|}{Sensitivity} \\
    \cline{2-4}
    & Normal & Pneumonia & COVID-19 \\
    \hline
    \hline
    COVID-Net~\cite{wang2020covid}  & 97.0 & 90.0 & 87.1 \\
    \hline
    Our Stage Pneumonia & \textbf{98.0} & \multicolumn{2}{c|}{90.84}  \\
    \hline
        Our Stage COVID-19 & - & 87.0  & \textbf{96.8}  \\
    \hline
    \end{tabular}%
  \label{tab.covidnetcomparison}%
  % \vspace{-3mm}
\end{table}%

%
% \section{Discussions}
% \label{sec:discuss}
\section{Conclusion}
\label{sec:conclusion}
Since the outbreak of COVID-19, intensive efforts have being made by healthcare experts in hospitals to reach diagnosis result for each patient even with supplemental symptoms. Suggested by medical experts, because of the efficiency and availability of CXR, we propose an AI-based screening system to recognize COVID-19 pneumonia in CXR images. Regarding coarse to fine manners, our cascaded method consists of lung segmentation, pneumonia recognition, and COVID-19 recognition as hierarchical screening. The proposed approach outperforms a previous method on open dataset of COVID-19 cases and is able to reach clinical-grade performance on NTUH and TMUH clinical data. Moreover, our method has been integrated into the internal system of Taiwan NHIA and CDC, achieving over 80\% sensitivity and specificity on NHIA clinical test cases. Key future research challenges include the sensitivity improvement on cases with mild symptoms, medical studies of COVID-19 cases with recognized lesion patterns, and the refinement of lung segmentation to uncover subtle and relevant regions.

% Since the outbreak of COVID-19, there are limited confirmed cases in Taiwan and only few doctors experienced in diagnosing SARS-CoV-2. It usually takes a couple days to reach a diagnosis result by doctors with cases showing no supplemental symptoms. To tackle this problem, NTUH healthcare experts suggested a centralized AI-assisted CXR screening solution. Instead of CT and other modalities, CXR is recommended because of its efficiency and wider availability.

\section{Acknowledgement}
\label{sec:acknowledgement}
We would like to thank National Taiwan University Hospital (NTUH), Taipei Medical University Hospital (TMUH), and Taiwan National Health Insurance Administration (NHIA) to provide clinical COVID-19 data for our studies. The experts from NTUH also provide us insightful feedback about medical findings and results evaluation on CXR images. For the deployment of our COVID-19 screening system, we appreciate the support from Taiwan Centers for Disease Control and Taiwan Executive Yuan, letting us contribute to the safeguard of Taiwan citizens' well-being during this global pandemic.

\bibliographystyle{unsrt}
\bibliography{covid-19}
\end{document}